\documentclass[sigconf, authorversion, screen]{acmart}

\setcopyright{rightsretained}
\acmDOI{10.1145/3287324.3293712}
\acmConference[SIGCSE '19]{the 50th ACM Technical Symposium on Computer Science Education}{February 27--March 2, 2019}{Minneapolis, MN, USA}
\acmYear{2019}
\copyrightyear{2019}
\acmSubmissionID{1040}

\begin{document}
\title{Subgoals, Problem Solving Phases, and Sources of Knowledge}
\subtitle{A Complex Mangle}

\author{Kevin Lin}
\affiliation{%
  \institution{University of California, Berkeley}
  \city{Berkeley}
  \state{California}
}
\email{kevinlin1@berkeley.edu}

\author{David DeLiema}
\affiliation{%
  \institution{University of California, Berkeley}
  \city{Berkeley}
  \state{California}
}
\email{deliema@berkeley.edu}

\begin{abstract}
Educational researchers have increasingly drawn attention to how students develop computational thinking (CT) skills, including in science, math, and literacy contexts. A key component of CT is the process of abstraction, a particularly challenging concept for novice programmers, but one vital to problem solving. We propose a framework based on situated cognition that can be used to document how instructors and students communicate about abstractions during the problem solving process. We develop this framework in a multimodal interaction analysis of a 32-minute long excerpt of a middle school student working in the PixelBots JavaScript programming environment at a two-week summer programming workshop taught by undergraduate CS majors. Through a microgenetic analysis of the process of teaching and learning about abstraction in this excerpt, we document the extemporaneous prioritization of subgoals and the back-and-forth coordination of problem solving phases. In our case study, we identify that (a) problem solving phases are nested with several instances of context-switching within a single phase; (b) the introduction of new ideas and information create bridges or opportunities to move between different problem solving phases; (c) planning to solve a problem is a non-linear process; and (d) pedagogical moves such as modeling and prompting highlight situated resources and advance problem solving. Future research should address how to help students structure subgoals and reflect on connections between problem solving phases, and how to help instructors reflect on their routes to supporting students in the problem solving process.
\end{abstract}



\maketitle

\section{Problem and Motivation}

Educational researchers have increasingly drawn attention to how students develop computational thinking (CT) skills \cite{Lye2014,Weintrop2015}, including in science, math, and literacy contexts \cite{Perez2018,Basu2017,Jacob2018}. A key component of CT is the process of \textit{abstraction}. Indeed, abstraction is recognized as a threshold concept \cite{Rountree2013}: a generative idea that once learned provides ``a qualitatively different view of subject matter within a discipline.'' The process of creating abstractions interleaves multiple problem solving phases: planning, building, and monitoring. In addition, creating abstractions requires attending to subgoals. Finally, pathways to learning about abstractions are structured by complex tools and driven by a variety of sources of knowledge, including perception, testimony, reasoning, and memory \cite{Chinn2011}. Our purpose in this paper is to stitch these elements into a framework that can be used to document how instructors and students communicate about computational thinking.

\section{Background and Related Work}

Our framework integrates constructs from the learning sciences, computer science educational research, and human computer interaction. Our point of departure is situated cognition, a theoretical framework that inextricably connects learning to interactions between tools, cognition, bodies, and communities of practice \cite{Jordan1995}. Increasingly, other researchers have studied CT teaching and learning from a situated perspective \cite{Lewis2012,Lewis2015,Flood2018Code,Flood2018Enskilment}. Specifically, our framework connects with prior work on how problem solving involves a balancing act between exploration, building, and monitoring \cite{Basu2017, Rountree2013}. In addition, we recognize that developing abstractions requires a gradual process of specifying and prioritizing goals and subgoals, including refactoring a previous route by developing new subgoals \cite{Basu2017}. The epistemic actions that advance problem solving are stretched across tools in the environment (e.g., editor, syntax checker, stepper tool, code reference sheet) and multiple sources of knowledge: perception, memory, reasoning, and testimony. Each interaction between a source of knowledge and a tool requires the learner to cross a gulf of execution, where they try to figure out how to express their idea to a tool in the environment, and attend to the outcome by crossing a gulf of evaluation, where they try to figure out the tool's response \cite{Norman2002}. Recognizing that causes of failure in programming emerge from complex connections between proximal and distal events \cite{Ko2005}, we assembled the framework above to comprehensively integrate multiple constructs and offer a new lens on the pathways students take to learn foundational computer programming concepts. 

\section{Approach and Uniqueness}

There is surprisingly little microgenetic, multimodal qualitative research on how young students program in naturalistic learning environments. For this paper, we selected a 32-minute long sample of a middle school student working in the PixelBots JavaScript programming environment at a two-week summer programming workshop taught by undergraduate CS majors. Using rich observational data including several camera angles and screen recordings, we transcribed the student's activities, repeatedly watched video, connected our observations to constructs noted above, and gradually developed our framework. This methodology blends interaction analysis with the constant comparative method \cite{Jordan1995,Glaser1965}. In this case study, the student solves the problem of programming the PixelBot to paint three jagged lines, horizontally spaced 3 tiles apart. Our purpose is to document the details of one student's problem solving process, rather than make generalized statements about learning processes. We hope that this research will allow for more rigorous experimental studies to come.

\section{Results and Contributions}

We identify the student's key challenge of coordinating subgoals and problem solving phases, and then identify how the student navigates this space by coordinating resources in the environment. 

\subsection{Coordinating Subgoals and Problem Solving Phases}

The focal student's programming process involves completing three subgoals---writing code to paint a \textit{jagged line}, writing code that \textit{moves to the next line}, and orchestrating this code in two functions---each of which entails navigating three phases of problem solving: planning, building, and monitoring. The student prioritizes subgoals and problem solving phases in response to syntax and logic bugs, instructor prompting, and focal sources of knowledge.


The student begins by \textit{planning} the trajectory of the PixelBot through the \textit{jagged line}, \textit{building} using only API movement instructions. She silently \textit{monitors} without running the code, and then deletes it, re-\textit{building} with painting actions and a loop. The student then attempts to \textit{monitor} the resulting PixelBot action by running the code, but quickly stops the program before it advances enough to show the corresponding PixelBot action. The student continues \textit{building} but a drag-and-drop attempt leads to an error message in the text editor. Her subsequent \textit{monitoring} and re-\textit{building} process, which generates more syntax bugs, involves trying to interpret error messages and create symmetry between brackets, parentheses, and quotes.

An instructor walks over and introduces a different monitoring process: comparing the broken syntax in the editor with correct syntax on a handout. Over six minutes, the student foregrounds the subgoal of writing the \textit{jagged line} code through three additional subgoals: \textit{writing the PixelBot trajectory}, \textit{adding color}, and \textit{adding a loop}. New information from the environment motivates the selection of subgoals. When errors are identified, the student takes immediate action rather than continuing with her current task. The student transitions rapidly between planning, building, and monitoring. In addition, within a single problem solving phase, we see a range of sources of knowledge deployed. For example, the student's monitoring approach of \textit{perceiving} error messages and \textit{reasoning} about symmetry contrasts with the instructor's monitoring approach of \textit{perceiving} and \textit{reasoning} about connections between the handout and editor. After resolving the syntax bugs, the student identifies the subgoal of developing the \textit{jagged line function} as complete despite having painted several squares the wrong color. Not until this logic bug is flagged in the correctness check at the end of the session does the student re-foreground the subgoal of \textit{monitoring} the jagged line function.

Throughout this process, the participants extemporaneously structure how to navigate the problem space, selectively managing subgoals. Their pathway is neither linear nor premeditated: planning arises throughout, contingent on syntax bugs that arise, logic bugs not yet noticed, moments of refactoring, and moments of recalibration after subgoals are judged complete. 

\subsection{Coordinating Resources}

How does the work of pursuing subgoals across problem solving phases unfold? This section describes a lower-level, moment-to-moment coordination of media across people and tasks.

The student utilizes multiple sources of knowledge to coordinate resources in the environment. For example, while monitoring the \textit{jagged line} code, the instructor foregrounds a process for syntax verification by comparing the code token-by-token with an example from a handout: ``Look at this [code reference sheet] and then compare every sort of word like `function', `function', [\ldots{}]'' The student looks back-and-forth between the screen and the handout, \textit{perceiving} and \textit{reasoning} about similarities between the two. These sources of knowledge thus bridge two resources in the environment: the handout and editor. This process is repeated as the student debugs a missing \verb$paint$ instruction while working on \textit{moving to the next line}. As the student steps line-by-line through the code, her gaze moves right and left between the code editor and the corresponding PixelBot action. Attention to different parts is variable: she steps quickly through parts she has already thoroughly vetted, and slows down, even stopping, when she arrives at code that corresponds with the dispreferred PixelBot action. The student again \textit{perceives} and \textit{reasons} about two resources (editor and PixelBot actions).

While monitoring the \textit{jagged line} code using the instructor's token-by-token syntax verification strategy, the student runs into a \verb$repeat$ loop which is not documented. The instructor provides expert \textit{testimony} and suggests clicking the button to insert the \verb$repeat$ template. The student coordinates the two snippets of \verb$repeat$ code in the editor: one serving as the template, and one serving as the target to be fixed, but applies the same process of token-by-token comparison as she did earlier with the code reference sheet. Later, while monitoring the code to \textit{move to the next line}, the student draws on her \textit{memory} and uses the same strategy of inserting the \verb$repeat$ template to check the syntax of the program. The student calls upon this strategy in two different instances to help cross the gulf of evaluation and propose a fix for each syntax bug.

The coordination of resources in the learning environment helps the student cross the gulf of evaluation and execution. Before the instructor foregrounds the syntax verification strategy, the student's method for resolving each syntax bug was to read the error message, interpret the problem, and propose a fix. This process presents a wide gulf of evaluation as the student needs to operationalize the error message by converting a description of the problem into an applicable fix based on prior knowledge of program syntax. The instructor's method for syntax verification relies on the same sources of knowledge, \textit{perception} and \textit{reasoning}, but uses the handout to present a more accessible affordance to cross the gulf of evaluation.

\subsection{Conclusion}
Through a microgenetic analysis of the process of teaching and learning about abstraction, we document the extemporaneous prioritization of subgoals and the back-and-forth coordination of problem solving phases. Participants navigate these tasks using multiple resources and sources of knowledge to cross gulfs of execution and evaluation. Future research should address how to help students structure subgoals, reflect on problem solving techniques, and recruit productive sources of knowledge as an ensemble process.

\begin{acks}
This material is based upon work supported by the \grantsponsor{SP784}{National Science Foundation}{https://doi.org/10.13039/100000001} under the Grant No.~\grantnum{SP784}{1607742},~\grantnum{SP784}{1612770}, and~\grantnum{SP784}{1612660}. Any opinions, findings, and conclusions or recommendations expressed in this material are those of the author(s) and do not necessarily reflect the views of the National Science Foundation.
\end{acks}

\bibliographystyle{ACM-Reference-Format}
\bibliography{bibliography}

\end{document}